\documentclass[twocolumn,letterpaper,aps,prc,longbibliography,superscriptaddress,showpacs,nofootinbib,floatfix]{revtex4-1}

\usepackage{graphicx}	
\usepackage{multirow}
\usepackage{amsmath}      
\usepackage{xspace}	

\newcommand{\pt}{\mbox{$p_T$}\xspace}

\newcommand{\sqsn}{\mbox{$\sqrt{s_{_{NN}}}$}\xspace}
\newcommand{\snn}{\mbox{$\sqrt{s_{_{NN}}}$}\xspace}

\newcommand{\pp}{\mbox{$p$$+$$p$}\xspace}
\renewcommand{\AA}{\mbox{$A$$+$$A$}\xspace}

\newcommand{\auau}{\mbox{Au$+$Au}\xspace}

\newcommand{\piz}{\mbox{$\pi^0$}\xspace}
\newcommand{\pizs}{\mbox{$\pi^0$}s\xspace}

\newcommand{\vtwo}{\mbox{$v_2$}\xspace}
\newcommand{\vthr}{\mbox{$v_3$}\xspace}

\newcommand{\gevc}{\mbox{GeV/$c$}\xspace}

\begin{document}

\title{Azimuthally anisotropic emission of low-momentum direct photons in 
Au$+$Au collisions at $\sqrt{s_{_{NN}}}=200$~GeV}


\newcommand{\abilene}{Abilene Christian University, Abilene, Texas 79699, USA}
\newcommand{\augie}{Department of Physics, Augustana University, Sioux Falls, South Dakota 57197, USA}
\newcommand{\banaras}{Department of Physics, Banaras Hindu University, Varanasi 221005, India}
\newcommand{\barc}{Bhabha Atomic Research Centre, Bombay 400 085, India}
\newcommand{\baruch}{Baruch College, City University of New York, New York, New York, 10010 USA}
\newcommand{\bnlcoll}{Collider-Accelerator Department, Brookhaven National Laboratory, Upton, New York 11973-5000, USA}
\newcommand{\bnlphys}{Physics Department, Brookhaven National Laboratory, Upton, New York 11973-5000, USA}
\newcommand{\caucr}{University of California-Riverside, Riverside, California 92521, USA}
\newcommand{\charlesczech}{Charles University, Ovocn\'{y} trh 5, Praha 1, 116 36, Prague, Czech Republic}
\newcommand{\chonbuk}{Chonbuk National University, Jeonju, 561-756, Korea}
\newcommand{\ciae}{Science and Technology on Nuclear Data Laboratory, China Institute of Atomic Energy, Beijing 102413, Peoples' Republic of~China}
\newcommand{\cns}{Center for Nuclear Study, Graduate School of Science, University of Tokyo, 7-3-1 Hongo, Bunkyo, Tokyo 113-0033, Japan}
\newcommand{\colorado}{University of Colorado, Boulder, Colorado 80309, USA}
\newcommand{\columbia}{Columbia University, New York, New York 10027 and Nevis Laboratories, Irvington, New York 10533, USA}
\newcommand{\czechtech}{Czech Technical University, Zikova 4, 166 36 Prague 6, Czech Republic}
\newcommand{\dapnia}{Dapnia, CEA Saclay, F-91191, Gif-sur-Yvette, France}
\newcommand{\debrecen}{Debrecen University, H-4010 Debrecen, Egyetem t{\'e}r 1, Hungary}
\newcommand{\elte}{ELTE, E{\"o}tv{\"o}s Lor{\'a}nd University, H-1117 Budapest, P{\'a}zm{\'a}ny P.~s.~1/A, Hungary}
\newcommand{\ewha}{Ewha Womans University, Seoul 120-750, Korea}
\newcommand{\fit}{Florida Institute of Technology, Melbourne, Florida 32901, USA}
\newcommand{\fsu}{Florida State University, Tallahassee, Florida 32306, USA}
\newcommand{\gsu}{Georgia State University, Atlanta, Georgia 30303, USA}
\newcommand{\hanyang}{Hanyang University, Seoul 133-792, Korea}
\newcommand{\hiroshima}{Hiroshima University, Kagamiyama, Higashi-Hiroshima 739-8526, Japan}
\newcommand{\howard}{Department of Physics and Astronomy, Howard University, Washington, DC 20059, USA}
\newcommand{\ihepprot}{IHEP Protvino, State Research Center of Russian Federation, Institute for High Energy Physics, Protvino, 142281, Russia}
\newcommand{\illuiuc}{University of Illinois at Urbana-Champaign, Urbana, Illinois 61801, USA}
\newcommand{\inrras}{Institute for Nuclear Research of the Russian Academy of Sciences, prospekt 60-letiya Oktyabrya 7a, Moscow 117312, Russia}
\newcommand{\instpasczech}{Institute of Physics, Academy of Sciences of the Czech Republic, Na Slovance 2, 182 21 Prague 8, Czech Republic}
\newcommand{\isu}{Iowa State University, Ames, Iowa 50011, USA}
\newcommand{\jaea}{Advanced Science Research Center, Japan Atomic Energy Agency, 2-4 Shirakata Shirane, Tokai-mura, Naka-gun, Ibaraki-ken 319-1195, Japan}
\newcommand{\jinrdubna}{Joint Institute for Nuclear Research, 141980 Dubna, Moscow Region, Russia}
\newcommand{\jyvaskyla}{Helsinki Institute of Physics and University of Jyv{\"a}skyl{\"a}, P.O.Box 35, FI-40014 Jyv{\"a}skyl{\"a}, Finland}
\newcommand{\karoly}{K\'aroly R\'oberts University College, H-3200 Gy\"ngy\"os, M\'atrai \'ut 36, Hungary}
\newcommand{\kek}{KEK, High Energy Accelerator Research Organization, Tsukuba, Ibaraki 305-0801, Japan}
\newcommand{\korea}{Korea University, Seoul, 136-701, Korea}
\newcommand{\kurchatov}{National Research Center ``Kurchatov Institute", Moscow, 123098 Russia}
\newcommand{\kyoto}{Kyoto University, Kyoto 606-8502, Japan}
\newcommand{\labllr}{Laboratoire Leprince-Ringuet, Ecole Polytechnique, CNRS-IN2P3, Route de Saclay, F-91128, Palaiseau, France}
\newcommand{\lahorelums}{Physics Department, Lahore University of Management Sciences, Lahore 54792, Pakistan}
\newcommand{\lawllnl}{Lawrence Livermore National Laboratory, Livermore, California 94550, USA}
\newcommand{\losalamos}{Los Alamos National Laboratory, Los Alamos, New Mexico 87545, USA}
\newcommand{\lpc}{LPC, Universit{\'e} Blaise Pascal, CNRS-IN2P3, Clermont-Fd, 63177 Aubiere Cedex, France}
\newcommand{\lund}{Department of Physics, Lund University, Box 118, SE-221 00 Lund, Sweden}
\newcommand{\maryland}{University of Maryland, College Park, Maryland 20742, USA}
\newcommand{\mass}{Department of Physics, University of Massachusetts, Amherst, Massachusetts 01003-9337, USA}
\newcommand{\michigan}{Department of Physics, University of Michigan, Ann Arbor, Michigan 48109-1040, USA}
\newcommand{\muenster}{Institut f\"ur Kernphysik, University of Muenster, D-48149 Muenster, Germany}
\newcommand{\muhlenberg}{Muhlenberg College, Allentown, Pennsylvania 18104-5586, USA}
\newcommand{\myongji}{Myongji University, Yongin, Kyonggido 449-728, Korea}
\newcommand{\nagasaki}{Nagasaki Institute of Applied Science, Nagasaki-shi, Nagasaki 851-0193, Japan}
\newcommand{\nara}{Nara Women's University, Kita-uoya Nishi-machi Nara 630-8506, Japan}
\newcommand{\natmephi}{National Research Nuclear University, MEPhI, Moscow Engineering Physics Institute, Moscow, 115409, Russia}
\newcommand{\newmex}{University of New Mexico, Albuquerque, New Mexico 87131, USA}
\newcommand{\nmsu}{New Mexico State University, Las Cruces, New Mexico 88003, USA}
\newcommand{\ohio}{Department of Physics and Astronomy, Ohio University, Athens, Ohio 45701, USA}
\newcommand{\ornl}{Oak Ridge National Laboratory, Oak Ridge, Tennessee 37831, USA}
\newcommand{\orsay}{IPN-Orsay, Univ.~Paris-Sud, CNRS/IN2P3, Universit\'e Paris-Saclay, BP1, F-91406, Orsay, France}
\newcommand{\peking}{Peking University, Beijing 100871, Peoples' Republic of~China}
\newcommand{\pnpi}{PNPI, Petersburg Nuclear Physics Institute, Gatchina, Leningrad region, 188300, Russia}
\newcommand{\riken}{RIKEN Nishina Center for Accelerator-Based Science, Wako, Saitama 351-0198, Japan}
\newcommand{\rikjrbrc}{RIKEN BNL Research Center, Brookhaven National Laboratory, Upton, New York 11973-5000, USA}
\newcommand{\rikkyo}{Physics Department, Rikkyo University, 3-34-1 Nishi-Ikebukuro, Toshima, Tokyo 171-8501, Japan}
\newcommand{\saispbstu}{Saint Petersburg State Polytechnic University, St.~Petersburg, 195251 Russia}
\newcommand{\saopaulo}{Universidade de S{\~a}o Paulo, Instituto de F\'{\i}sica, Caixa Postal 66318, S{\~a}o Paulo CEP05315-970, Brazil}
\newcommand{\seoulnat}{Department of Physics and Astronomy, Seoul National University, Seoul 151-742, Korea}
\newcommand{\stonybrkc}{Chemistry Department, Stony Brook University, SUNY, Stony Brook, New York 11794-3400, USA}
\newcommand{\stonycrkp}{Department of Physics and Astronomy, Stony Brook University, SUNY, Stony Brook, New York 11794-3800, USA}
\newcommand{\sungskku}{Accelerator and Medical Instrumentation Engineering Lab, SungKyunKwan University, 53 Myeongnyun-dong, 3-ga, Jongno-gu, Seoul 110-745, Korea}
\newcommand{\tenn}{University of Tennessee, Knoxville, Tennessee 37996, USA}
\newcommand{\titech}{Department of Physics, Tokyo Institute of Technology, Oh-okayama, Meguro, Tokyo 152-8551, Japan}
\newcommand{\tsukuba}{Center for Integrated Research in Fundamental Science and Engineering, University of Tsukuba, Tsukuba, Ibaraki 305, Japan}
\newcommand{\vandy}{Vanderbilt University, Nashville, Tennessee 37235, USA}
\newcommand{\waseda}{Waseda University, Advanced Research Institute for Science and Engineering, 17  Kikui-cho, Shinjuku-ku, Tokyo 162-0044, Japan}
\newcommand{\weizmann}{Weizmann Institute, Rehovot 76100, Israel}
\newcommand{\wigner}{Institute for Particle and Nuclear Physics, Wigner Research Centre for Physics, Hungarian Academy of Sciences (Wigner RCP, RMKI) H-1525 Budapest 114, POBox 49, Budapest, Hungary}
\newcommand{\yonsei}{Yonsei University, IPAP, Seoul 120-749, Korea}
\newcommand{\zagreb}{University of Zagreb, Faculty of Science, Department of Physics, Bijeni\v{c}ka 32, HR-10002 Zagreb, Croatia}
\affiliation{\abilene}
\affiliation{\augie}
\affiliation{\banaras}
\affiliation{\barc}
\affiliation{\baruch}
\affiliation{\bnlcoll}
\affiliation{\bnlphys}
\affiliation{\caucr}
\affiliation{\charlesczech}
\affiliation{\chonbuk}
\affiliation{\ciae}
\affiliation{\cns}
\affiliation{\colorado}
\affiliation{\columbia}
\affiliation{\czechtech}
\affiliation{\dapnia}
\affiliation{\debrecen}
\affiliation{\elte}
\affiliation{\ewha}
\affiliation{\fit}
\affiliation{\fsu}
\affiliation{\gsu}
\affiliation{\hanyang}
\affiliation{\hiroshima}
\affiliation{\howard}
\affiliation{\ihepprot}
\affiliation{\illuiuc}
\affiliation{\inrras}
\affiliation{\instpasczech}
\affiliation{\isu}
\affiliation{\jaea}
\affiliation{\jinrdubna}
\affiliation{\jyvaskyla}
\affiliation{\karoly}
\affiliation{\kek}
\affiliation{\korea}
\affiliation{\kurchatov}
\affiliation{\kyoto}
\affiliation{\labllr}
\affiliation{\lahorelums}
\affiliation{\lawllnl}
\affiliation{\losalamos}
\affiliation{\lpc}
\affiliation{\lund}
\affiliation{\maryland}
\affiliation{\mass}
\affiliation{\michigan}
\affiliation{\muenster}
\affiliation{\muhlenberg}
\affiliation{\myongji}
\affiliation{\nagasaki}
\affiliation{\nara}
\affiliation{\natmephi}
\affiliation{\newmex}
\affiliation{\nmsu}
\affiliation{\ohio}
\affiliation{\ornl}
\affiliation{\orsay}
\affiliation{\peking}
\affiliation{\pnpi}
\affiliation{\riken}
\affiliation{\rikjrbrc}
\affiliation{\rikkyo}
\affiliation{\saispbstu}
\affiliation{\saopaulo}
\affiliation{\seoulnat}
\affiliation{\stonybrkc}
\affiliation{\stonycrkp}
\affiliation{\sungskku}
\affiliation{\tenn}
\affiliation{\titech}
\affiliation{\tsukuba}
\affiliation{\vandy}
\affiliation{\waseda}
\affiliation{\weizmann}
\affiliation{\wigner}
\affiliation{\yonsei}
\affiliation{\zagreb}
\author{A.~Adare} \affiliation{\colorado} 
\author{S.~Afanasiev} \affiliation{\jinrdubna} 
\author{C.~Aidala} \affiliation{\losalamos} \affiliation{\mass} \affiliation{\michigan} 
\author{N.N.~Ajitanand} \affiliation{\stonybrkc} 
\author{Y.~Akiba} \affiliation{\riken} \affiliation{\rikjrbrc} 
\author{R.~Akimoto} \affiliation{\cns} 
\author{H.~Al-Bataineh} \affiliation{\nmsu} 
\author{J.~Alexander} \affiliation{\stonybrkc} 
\author{M.~Alfred} \affiliation{\howard} 
\author{H.~Al-Ta'ani} \affiliation{\nmsu} 
% ...
\collaboration{PHENIX Collaboration} \noaffiliation

\date{\today}

\begin{abstract}


The PHENIX experiment at the Relativistic Heavy Ion Collider has measured 
2nd and 3rd order Fourier coefficients of the azimuthal distributions of 
direct photons emitted at midrapidity in Au$+$Au collisions at 
$\sqrt{s_{_{NN}}}=200$~GeV for various collision centralities. Combining 
two different analysis techniques, results were obtained in the transverse 
momentum range of $0.4<p_{T}<4.0$~GeV/$c$.  At low $p_T$ the 
second-order coefficients, $v_2$, are similar to the ones observed in 
hadrons. Third order coefficients, $v_3$, are nonzero and almost 
independent of centrality.  These new results on $v_2$ and $v_3$, combined 
with previously published results on yields, are compared to model 
calculations that provide yields and asymmetries in the same framework. 
Those models are challenged to explain simultaneously the observed large 
yield and large azimuthal anisotropies.

\end{abstract}

\pacs{25.75.Dw} 
	
\keywords{fractional momentum loss}

   \maketitle

		\section{Introduction}

Direct photons emerging from relativistic heavy ion collisions have long 
been considered an important probe of the entire evolution of the 
colliding system~\cite{Shuryak:1978ph}.  At almost all known or 
conjectured stages of the collision there are processes producing photons. 
Unlike hadronic observables that mostly encode the state of the medium at 
 freeze-out, photons are emitted at all times throughout the rapid 
evolution of the heavy ion collision and leave the interaction region 
unmodified. Thus by measuring direct photons one has access to information 
about the properties and dynamics of the medium integrated over space and 
time. The measurement of direct photons is challenging due to a large 
background of photons from the vacuum decay of final state hadrons (\piz, 
$\eta$, $\omega$, etc.).
 
The PHENIX experiment at the Relativistic Heavy Ion Collider reported 
large direct photon yields~\cite{ppg086} with strong 
centrality dependence~\cite{ppg162} and significant azimuthal anisotropy 
or ``elliptic flow''~\cite{ppg126}. Particularly surprising is the 
discovery of large azimuthal anisotropy for direct 
photons~\cite{ppg126}, which is comparable to that observed for 
hadrons~\cite{ppg132}.  Preliminary results from the Large Hadron 
Collider~\cite{wilde:2013phot,lohner:2013alicev2} indicate similar 
direct photon yields and anisotropies.  
The observation of large azimuthal anisotropy combined with observations 
published earlier that the direct photon yields themselves are 
large~\cite{ppg086,ppg162} contradicts several existing interpretations 
where the large yields are provided at the very early production stage, 
when the temperature of the system is highest but the collective flow 
including azimuthal asymmetry is negligible. Conversely, the observed 
large anisotropy suggests that photon production occurs at very late 
stages of the collision when the collective flow of the system is fully 
developed, while the temperature and the corresponding thermal photon 
emission rates are already lower. Indeed, theoretical models have great 
difficulty to simultaneously describe the observed yields and 
anisotropy. This failure, colloquially called ``the direct photon 
puzzle'', triggered a large amount of theoretical work, new models and 
insights~\cite{turbide:2006flow,chatterjee:2006flow,chatterjee:2009flow,dusling:2010flow,hees:2011fb,linnyk:phsd2014,gale:2015semi,Muller:2013ila,shen:2013vja,shen:2015visc,hees:2015pce,monnai:2014scf,chatterjee:2013flow,dion:2011visc,fmliu:2014ft,vujanovic:2014early,mclerran:2014glas,tuchin:2013magn,mclerran:2015tail,basar:2012conf,gelis:2004noneq,biro:2014illu,pisarski:2015poly,campbell}.

In this paper we present new, more precise results on the azimuthal 
anisotropy of direct photon emission from 200\,GeV \auau collisions 
recorded in 2007 and 2010 by the PHENIX experiment. Results include second 
and third order Fourier components of azimuthal distributions (\vtwo and 
\vthr, respectively) measured over a transverse momentum range extended 
down to 0.4\,\gevc. The new data, together with published results on 
yields, are compared to some of the more recent model calculations.

The paper is organized as follows.  In Sec.~\ref{sec:expgeneral} we 
describe the experiment, the data set, the way events are selected and 
categorized, and the two methods by which photons are measured.  In 
Sec.~\ref{sec:expv2} the steps needed to determine the direct photon 
\vtwo, \vthr, and their uncertainties are described, and the final results 
are presented.  In Sec.~\ref{sec:modelcomp} the results are compared to a 
few models treating yields and azimuthal asymmetries in a consistent 
framework.  Sec.~\ref{sec:summary} summarizes our findings.

		\section{Experimental setup and photon measurements}
		\label{sec:expgeneral}

In PHENIX photons are detected by two substantially different techniques. 
The first technique uses external conversion of photons as described in 
detail in Ref.~\cite{ppg162}.  This method provides a high purity photon 
sample with good momentum resolution, but requires large statistics due to 
the few percent conversion probability and reduced acceptance. Therefore 
the \pt range is limited. The second technique is a traditional 
calorimetric measurement of photons similar to Ref.~\cite{ppg126}, but 
with higher statistics. For photons identified by either technique, the 
azimuthal anisotropy is extracted with the event plane (EP) method. Here 
we give a brief summary of the PHENIX detector systems and a short 
description of the two analyses.

	\subsection{Event selection and centrality determination}

Data from 200\,GeV \auau collisions were recorded with a minimum bias (MB) 
trigger based on the signal in the beam-beam counters (BBC)~\cite{nimbbc}, 
which are located around the beampipe at $3.1<|\eta|<3.9$ and cover the 
full azimuth. The minimum bias trigger requires at least two hits in each 
of the two BBCs (north and south) as well as a reconstructed vertex from 
the time-of-flight difference between the two sides.  The efficiency of 
the MB trigger is $92.3\pm0.4(stat)\pm1.6(sys)$\%.

Collision centrality is calculated as percentiles of the total charge 
distribution in the north and south BBC. The centrality determination is 
based on percentiles of the total charge seen in the north and south BBC 
and takes into account small shifts in $\eta$ coverage due to variations 
of the collision $z$-vertex.

	\subsection{Inclusive photons via external conversion}
	\label{sec:Conv_photons}

External conversion photons are reconstructed from $2.6\times10^9$ MB \sqsn = 200\,GeV \auau events recorded during the 2010 data taking 
period. The event vertex in this dataset was $|z|<10$\,cm to ensure that 
the magnetic field is sufficiently uniform. The same sample was previously 
used in Ref.~\cite{ppg162} to determine direct photon yield and its 
centrality dependence, where details of this analysis can be found. In the 
rest of this paper this sample is referred to as ``conversion photons''.

Photons convert to $e^{+}e^{-}$ pairs in the readout plane of the Hadron 
Blind Detector (HBD)~\cite{nimhbd}, which is located 
at $\sim$60\,cm 
radial distance from the collision vertex and corresponds to $\sim$3\% 
$X_0$, where $X_{0}$ is the radiation length. The electron and positron 
from the photon conversion are tracked through the PHENIX central tracking 
detectors~\cite{nimtrk}. The azimuthal direction $\phi$ and the momentum 
$p$ are reconstructed from the drift-chamber information, while the 
polar angle of each track is determined by a point measurement in the 
innermost pad-chamber (PC1) and the collision vertex. High efficiency 
electron identification cuts are used to reduce the hadron contamination 
in the sample. Light above a minimum threshold in the ring-imaging 
\v{C}erenkov detector~\cite{nimrich} and a matching cluster of energy 
$E$ in the electromagnetic calorimeter (EMCal)~\cite{nimemc} such that 
$E>0.15$\,GeV and $E/p>0.5$, where $p$ is the momentum, are required. The 
EMCal comprises two calorimeter types: 6 sectors of lead scintillator 
sampling calorimeter (PbSc) and 2 sectors of lead glass \v{C}erenkov 
calorimeter (PbGl). The typical energy resolution of the PbSc is $\delta 
E/E=8.1\,\%/\sqrt{E(GeV)}\oplus2.1\,\%$, and that of the PbGl is $\delta 
E/E=5.9\,\%/\sqrt{E(GeV)}\oplus0.8\,\%$.
The energy resolution, just as the photon identification efficiency,
depends on centrality and its (small) effect is corrected for using simulated 
photon showers embedded into real events.

All remaining tracks with \pt$>0.2$\,\gevc, are combined to pairs. 
Conversion photons are identified by analyzing the invariant mass of the 
pairs. The default tracking in PHENIX assumes that each track originates 
at the collision vertex. Thus, if the $e^{+}e^{-}$ pair comes from a 
conversion of a real photon in the HBD readout plane, the momenta will be 
mis-measured and a finite mass, in this case about 
$m_{ee}\sim$12\,MeV/$c^2$, is reconstructed. Conversely, if the momenta 
are re-calculated assuming the HBD readout plane as origin, the invariant 
mass is close to zero. Through a simultaneous cut on both mass 
calculations a sample of photon conversions with a purity of 99\% is 
obtained down to \pt = 0.4\,\gevc~\cite{ppg162}. The remaining 1\% of 
pairs are mostly from the \piz Dalitz decays. The effect on the inclusive photon $v_n$ is estimated to be smaller than 1\%. 
  
	\subsection{Inclusive photons and \pizs via the calorimeter}
	\label{sec:EMCAL_photons}

The PHENIX EMCal is the principal detector in the calorimetric analysis, 
which is performed in a similar way as in Ref.~\cite{ppg126}.  The \vtwo 
and \vthr are measured simultaneously for inclusive photons and \pizs.  A 
total of $4.4\times10^{9}$ MB \auau events from the 2007 data taking 
period are analyzed.  The event vertex in this sample was $|z|<30$\,cm.

Photon candidates in the EMCal are clusters above a threshold energy of 
0.2\,GeV that pass a shower shape cut as well as a charged particle veto 
cut by the pad chamber PC3 immediately in front of the EMCal. 
However, 
photon candidates with less than 1\, GeV energy are only used to reconstruct
\piz, but are not included in the inclusive photon sample of the calorimeter.
As described in Ref.~\cite{Afanasiev:2012dg}, the remaining hadron 
contamination was estimated by comparing {\sc geant} simulations, 
verified with actual data.
The \piz is 
measured via the 2$\gamma$ decay channel, with a cut on the energy 
asymmetry of the two photons $\alpha = \frac{|E_1-E_2|}{E_1+E_2} <$ 0.8.  
For each \pt bin the number of reconstructed \pizs is taken as the 
integral of the two-photon invariant mass distribution, with the 
combinatorial background subtracted by the mixed event 
method~\cite{ppg080}.  The signal to background ratio at 
$1.0<p_{T}<1.5$\,\gevc is 0.1, rapidly improving with increasing \pt.

For the inclusive photon measurement it is important to restrict the 
measurement to a region where the residual contamination from 
misidentified hadrons is small.  Therefore, in the inclusive photon 
sample only clusters with $E>1$\,GeV are considered. On the other 
hand the inclusive (and direct) photon results presented here have 
an upper range of 4~GeV/$c$, which is far from the threshold where 
two decay photons from a $\pi^0$ can merge in the calorimeter.  
Within this \pt range a purity of larger than 95\% is achieved. 
The largest contamination of the photon sample results from 
antineutrons, which are not removed by the charge particle veto
but deposit significant energy through annihilation. 
The systematic uncertainty from 
particle identification (PID) of photons is estimated by varying 
both the shower shape cut (five different settings) and, 
independently, applying or omitting the charged particle veto cut.  
Results from all cut variations are then fully corrected. The deviation 
between results is 3-4\%, which is quoted as systematic uncertainty on the 
inclusive photon yield. 

	\subsection{Event plane determination}
	\label{sec:EP}

\begin{figure}[htbp]
\includegraphics[width=0.998\linewidth]{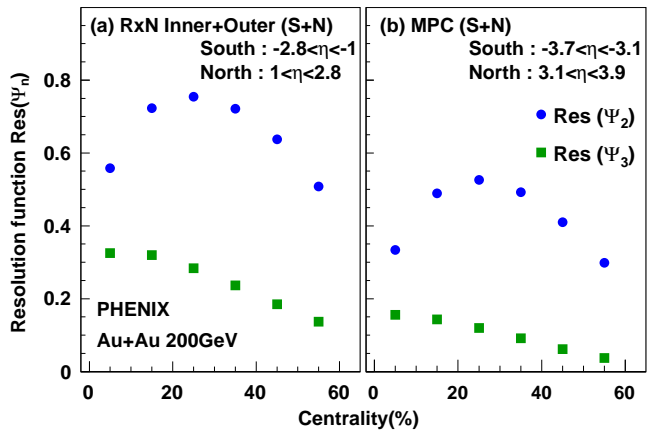}
  \caption{Event plane resolution as a function of centrality for the RxN(I+O) 
detector (a) used for the final results in this paper, and (b) for the MPC 
detector used to cross-check the results.
  }
    \label{fig:eventplanereso}
\end{figure}

PHENIX has different detector systems to establish the EP, which cover 
different pseudorapidity ($\eta$) ranges:  the outer and inner 
reaction plane 
detector (RxNO, 1$<|\eta|<$1.5, RxNI, 1.5$<|\eta|<$2.8), the muon piston 
calorimeters (MPCS, -3.7$<\eta<$-3.1, MPCN, 3.1$<\eta<$3.9), and the BBC 
(3.1$<|\eta|<$3.9).  All these detectors cover the full $2\pi$ azimuth 
and are sufficiently separated in $\eta$ such that we do not 
expect auto correlations between the event plane determination and the 
photon production asymmetry measured.  The RxNI and RXNO are 
scintillation counter systems with a 2\,cm Pb converter that makes them 
sensitive to photons in addition to charged particles.  
While these photons contribute to the determination of the event
plane, note that they are separated at least $\Delta\eta$=0.7
from the central region, which is where the 
photon $v_2$ and $v_3$ are measured.

The results in this paper are obtained using the event planes measured by 
the combination of the RxNI and RxNO~\cite{nimrxn}. Due to the large 
rapidity coverage this combination has the best resolution. The resolution 
$Res(\Psi_n)$ is measured with the 2-subevent 
method~\cite{poskanzer:1998sube}.  The resolution for RxN and MPC is shown 
in Figure~\ref{fig:eventplanereso}. The final results are cross-checked by 
using the other detectors for the event plane determination. Despite the 
significant difference in resolution the measured direct photon 
anisotropies are consistent, within the systematic uncertainties.

\begin{figure*}[htbp]
  \includegraphics[width=1.0\linewidth]{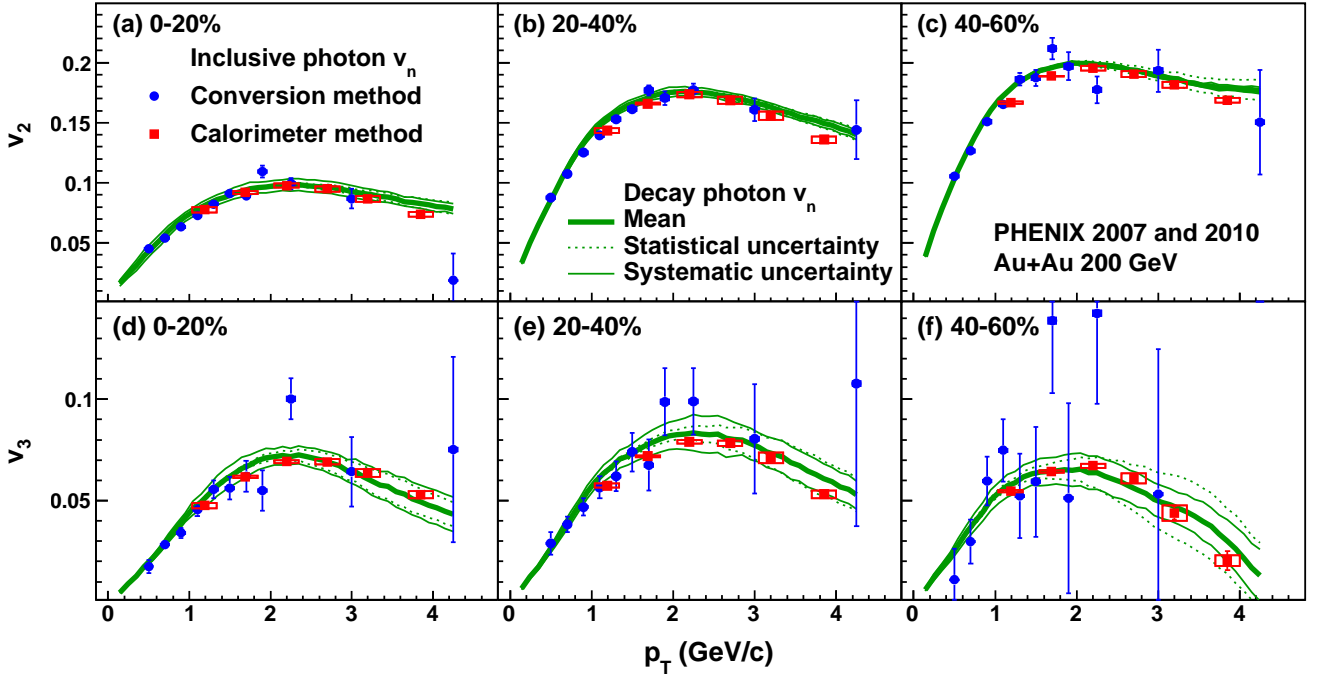}
  \caption{Inclusive photon \vtwo and \vthr at midrapidity ($|\eta| <$ 0.35) for 
\auau collisions at \snn = 200 GeV in different centrality bins 0\%--20\% 
(a,d), 20\%--40\% (b,e), and 40\%--60\% (c,f) with the event plane 
estimated with the reaction plane detector (1$<|\eta|<$2.8). The data from 
the external conversion method are shown as solid circles and from the 
calorimeter method as solid squares. The error bars (boxes) around the 
data points are statistical (systematic) uncertainties. Also shown are the 
calculated decay photon \vtwo and \vthr (thick solid line) along with the 
statistical (dotted line) and systematic (light solid line) uncertainties 
resulting from uncertainties on the input data. An additional systematic 
uncertainty due to the finite event plane resolution is not shown (see 
Table~\ref{tab:syserr}), because it is common to all $v_n$ measurements.
  }
    \label{fig:vincl}
\end{figure*}

		\section{Direct photon {\vtwo} and {\vthr}}
		\label{sec:expv2}

The photon anisotropy is measured via the coefficients of a Fourier 
decomposition of the azimuthal distributions of photons with respect to 
the event plane~\cite{poskanzer:1998sube}

\begin{equation}
\frac{dN}{d(\phi-\Psi_k)} \propto 1+ 
\sum\limits_{n} [v_{kn} \cos{ \{ n(\phi-\Psi_k)\} }],
\label{eq:flow}
\end{equation}

\noindent
where $\phi$ is the azimuthal angle of the photon, $\Psi_k$ is the 
orientation of the $k^{th}$ event plane for a given event, and $v_{kn}$ 
are the $n^{th}$ coefficients with respect to the $k^{th}$ event plane. In 
our analysis we made and explicitly tested the assumption that the 2nd 
and 3rd order event planes are uncorrelated, which allows us to ignore the 
$k\neq n$ terms and to introduce the notation \vtwo and \vthr for the case 
$k=n$, i.e. in the rest of the paper we use $v_2 \equiv v_{22}$ and $v_3 
\equiv v_{33}$.

The determination of the direct photon \vtwo and \vthr proceeds in three 
steps: (i) \vtwo and \vthr are determined for the conversion photon sample 
(Section~\ref{sec:Conv_photons}) and for the calorimeter photon sample 
(Section~\ref{sec:EMCAL_photons}) with respect to the event plane 
(Section~\ref{sec:EP}). We refer to these coefficients as inclusive photon 
$v^{\rm inc}_{2}$ and $v^{\rm inc}_3$. In the second step (ii), the decay photon 
$v^{\rm dec}_2$ and $v^{\rm dec}_3$ are estimated, i.e. the anisotropy resulting 
from the decays of hadrons to photons. It is calculated based on \vtwo, 
\vthr, and yields measured for charged and neutral pions; contributions 
from heavier mesons are taken into account using proper scaling (see 
Sec.~\ref{sec:decv2v3}). As a final step (iii) the direct photon \vtwo and 
\vthr are calculated statistically through a subtraction of the results 
from step (i) and (ii) weighted by the ratio $R_\gamma$, the ratio of the 
yields of direct photons to the yield of photons from hadron decays (see 
Eq.~\ref{eq:vdirect}).

	\subsection{Inclusive photon \vtwo and \vthr}
	\label{sec:incv2v3}

The inclusive photon \vtwo and \vthr are measured with respect to the 
event plane.  We employ two methods to determine these coefficients. For 
each photon the azimuthal angular difference $(\phi-\Psi_k)$, with 
$k=2,3$, is calculated. In the first method the coefficients are 
determined as the event ensemble average for individual bins in photon \pt 
and centrality:

\begin{equation}
v_{n} = \langle \cos{\{ n (\phi-\Psi_n)}\} \rangle/Res(\Psi_n). 
\label{eq:avgcos}
\end{equation}

\noindent
Here $Res(\Psi_n)$ is the resolution function that accounts for the finite 
event plane resolution (see Figure~\ref{fig:eventplanereso}).

In the second method the azimuthal distribution of photons in a given \pt 
and centrality bin is fitted as:

\begin{equation}
\frac{dN}{d(\phi-\Psi_n)} = N_0 [1+ 2 v'_{n} \cos{\{ n (\phi-\Psi_n)\} }], 
\label{eq:flowfit}
\end{equation}
\begin{equation}
v_n = v'_{n}/Res(\Psi_n).
\end{equation}

This is Eq.~\ref{eq:flow} for the case $k=n$ and neglecting all 
$k\neq n$ terms. The measured values of \vtwo and \vthr ($v'_2, v'_3$) 
need to be corrected for the event plane resolution.

In the conversion photon method the quoted $v_n$ values come from the 
average cosine method, while in the calorimeter analysis the quoted $v_n$ 
values are the average of the results obtained with the two methods. The 
difference between the two methods is less than 1\%. The results 
for the inclusive photon \vtwo and \vthr are shown in 
Figure~\ref{fig:vincl}. Both measurements agree in the region where they 
overlap.

	\subsection{Decay photon \vtwo and \vthr}
	\label{sec:decv2v3}

\begin{figure}[htbp]
\includegraphics[width=0.998\linewidth]{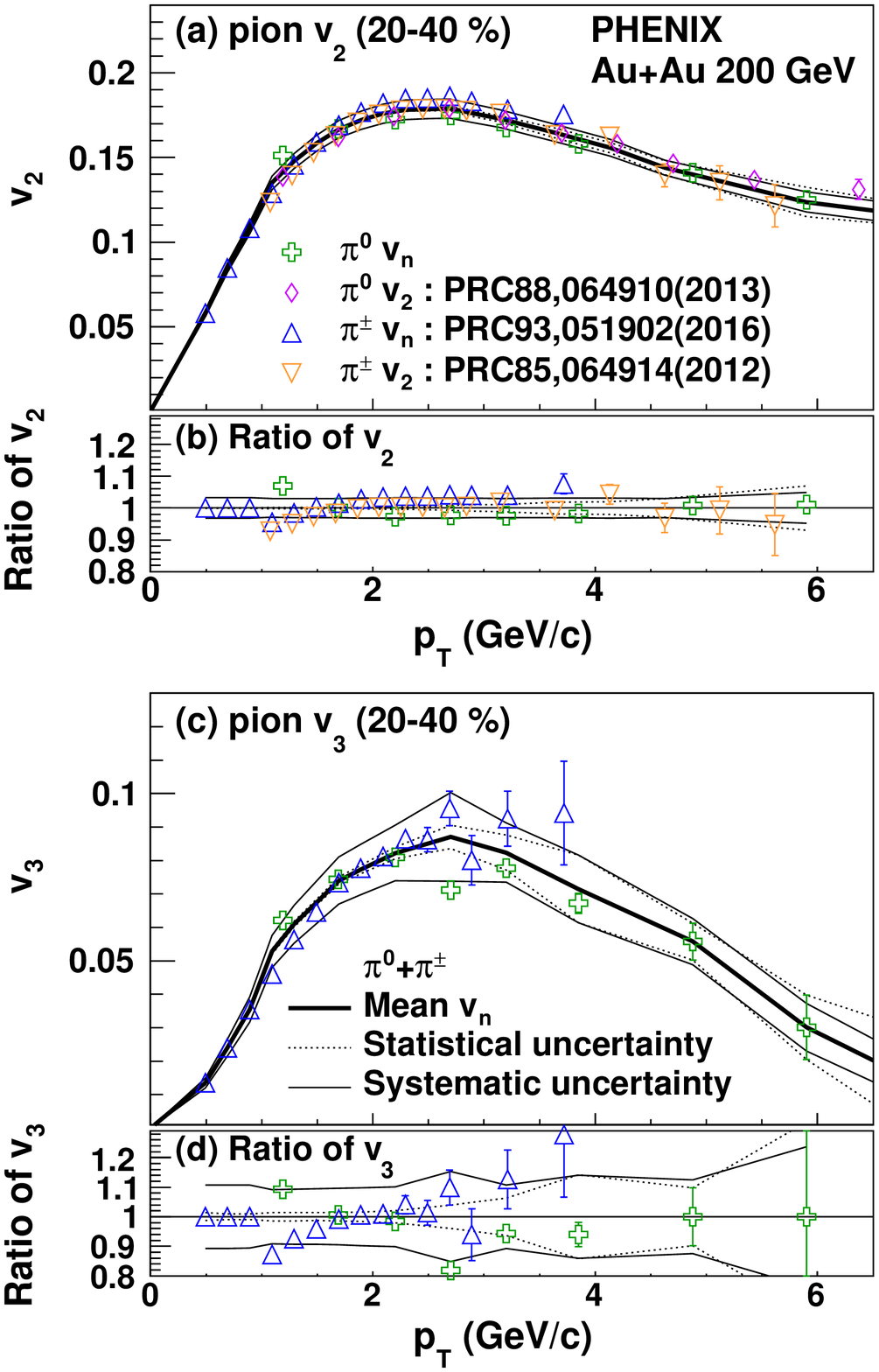}
\caption{Top panels: charged and neutral pion \vtwo (a) and \vthr (c) for the 
20\%--40\% centrality class, including previously published results. The 
averaged values used in our analysis are shown as a thick solid line 
together with the estimated statistical (dotted line) and systematic 
(light solid line) uncertainties. Bottom panels: ratio of the measured 
\vtwo (b) and \vthr (d) values to the averaged values. }
\label{fig:vpion}
\end{figure}

About 80\%--90\% of the inclusive photons come from decays of neutral mesons 
and exhibit an anisotropy with respect to the event plane that results 
from the anisotropy of the parent mesons~\cite{ppg126}. To estimate this 
contribution we use measured yields and anisotropy for charged and neutral 
pions; $v_n$ for heavier mesons is obtained by $KE_T$ scaling as described 
below. The yields of mesons used here are the same as are used for the 
measurement of $R_{\gamma}$ in Ref.~\cite{ppg162}.

The \vtwo and \vthr for pions are determined by combining data from 
different measurements of charged and neutral pion \vtwo and \vthr. The 
\piz \vtwo has been published in Ref.~\cite{ppg129} but the measurement 
has been repeated in this analysis to check the consistency of the 
results.  The method to count the number of \pizs in any \pt bin is 
briefly described in Sec.~\ref{sec:EMCAL_photons}. To obtain 
\vtwo (\vthr) for each \pt the number of reconstructed \pizs is extracted 
in six 15 (10) degree wide bins of the azimuthal angle $\Delta\Phi = \Phi 
- \Psi_n$ where $\Phi$ is the azimuth of the \piz and $\Psi_n$ is the 
second (third) order event plane (see Sec.~\ref{sec:EP}).  These 
distributions of the raw \piz counts vs.~$\Delta\Phi$ are then fitted as 
described in Sec.~\ref{sec:incv2v3} to obtain \vtwo and \vthr for \piz.  
Note that because the individual \pizs are not identified, the average 
cosine method~\cite{poskanzer:1998sube} is not applicable.

These data are combined with $\pi^{\pm}$ data in the \pt range 0.5 to 4 
\gevc~\cite{ppg147}. For \vtwo we also use $\pi^{\pm}$ data from 
Ref.~\cite{ppg123}. For the centrality class 20\%--40\% these data are 
compiled in Figure~\ref{fig:vpion}.  We interpolate the data, weighted by 
their statistical and systematic uncertainties, to obtain an average value 
$v_n$ for pions as a function of \pt. The result of this averaging 
procedure, including our estimate of the systematic uncertainties, is also 
shown in Figure~\ref{fig:vpion}.

For the heavier mesons, $\eta, \omega, \rho, \eta^{'}$, the $v_n$ is 
derived from the $v_n$ of the pions by scaling with the kinetic energy 
\cite{ppg062,ppg147}.

\begin{equation}
v^{meson}_{n} (KE_{T}) = v^{\pi}_n (KE_T),
\label{eq:ketscaling}
\end{equation}
\noindent where 
\begin{equation}
KE_T = m_T - m = \sqrt{p_T^2+m^2} - m,
\end{equation}
where $m$ is the mass of the corresponding meson. 

The yields of the heavier mesons are determined from the \piz yields at 
\pt = 5\,\gevc using the following ratios: $\eta/\pi^0 = 0.46\pm0.060$, 
$\omega/\pi^0 = 0.83\pm0.12$, $\rho/\pi^0 = 1.00\pm0.300$ and 
$\eta^{'}/\pi^0 = 0.25\pm0.075$. 
Below \pt = 2\,\gevc $KE_{T}$-scaling is only an extrapolation
for the $\eta$ yields.  Therefore, we also applied a
blast-wave fit, and the difference is included in the systematic
uncertainties. Note that the blast-wave fit results in lower $\eta$
yield at small \pt, increasing the direct photon yield and its \vtwo,\vthr.
The meson yields, momentum spectra and 
$v_n$ are used to simulate mesons that are then decayed to all decay 
chains including photons. From the simulation we calculate the decay 
photon $v_n^{\rm dec}$ using Eq.~\ref{eq:avgcos} with $Res(\Psi_n)=1$, 
because the event plane is known in the simulation. The only source of 
systematic uncertainty on $v_n^{\rm dec}$ is the uncertainty of the measured 
\piz \vtwo and \vthr, and the resulting decay photon \vtwo and \vthr, derived from it.  
The resulting $v_n^{\rm dec}$ is compared to the inclusive photon $v_n$ in 
Figure~\ref{fig:vincl}. We find that the decay photon and inclusive photon 
$v_n$ are similar. This was already observed for \vtwo in 
Ref.~\cite{ppg126}, but is now also found for \vthr. Given that a finite 
direct photon yield has already been established~\cite{ppg086,ppg162}, the 
similarity of $v_3^{\rm inc}$ and $v_3^{\rm dec}$ implies a large direct photon 
\vthr, as will be shown in the next section.

	\subsection{Direct photon \vtwo and \vthr}
	\label{sec:dirv2v3}

\begin{figure}[htbp]
\includegraphics[width=0.998\linewidth]{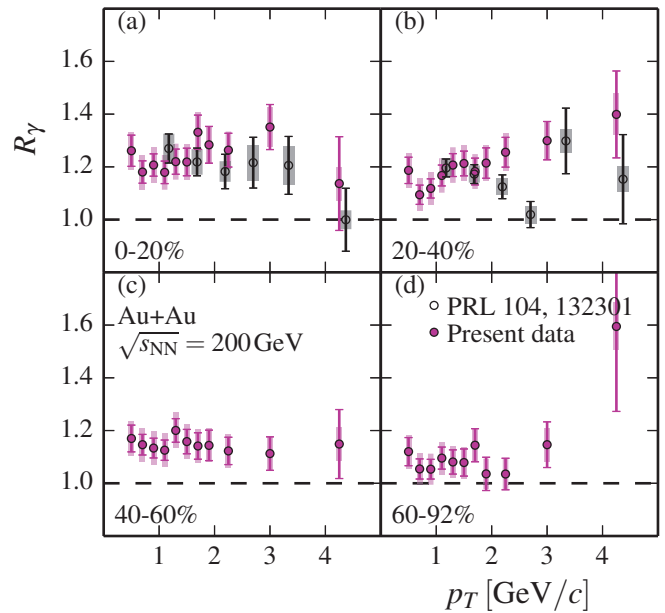}
\caption{The inclusive over decay photon ratio $R_{\gamma}$ 
used in the current analysis.  Present data means the results 
published in Ref.~\cite{ppg162}.}
\label{fig:rgamma}
\end{figure}

The \vtwo and \vthr for direct photons are extracted from the measured 
inclusive photon $v^{\rm inc}_n$, the decay photon $v_n^{\rm dec}$, discussed in 
the previous sections, and the ratio of the inclusive to decay photon 
yield $R_\gamma$ measured in Ref.~\cite{ppg162}. The procedure was 
introduced in Ref.~\cite{ppg126}:
\begin{equation}
v_n^{\rm dir} = \frac{R_{\gamma}v_n^{\rm inc} - v_n^{\rm dec}}{R_{\gamma}-1}.
\label{eq:vdirect}
\end{equation}
We reproduce $R_\gamma$ from Ref.~\cite{ppg162} with statistical and 
systematic uncertainties in Figure~\ref{fig:rgamma}.

All systematic uncertainties on the individual contributions on 
$v_n^{\rm dir}$ are summarized in Table~\ref{tab:syserr}. Uncertainties that 
are uncorrelated between data points are called Type A, those that are 
correlated are Type B and uncertainties that change all points by a common 
multiplicative factor are called Type C. Uncertainties on $R_\gamma$ are 
common for \vtwo and \vthr and for the conversion and calorimeter method. 
For photon and pion $v_n$ measurements with PHENIX the orientation of the 
event planes, i.e. $\Psi_n$, is determined with the same detectors using 
the same algorithms. Thus the systematic uncertainty on the event plane 
determination is common for all \vtwo (\vthr) measurements. The 
uncertainties on the decay photon $v_n$ are common to the conversion and 
calorimeter method. The systematic uncertainty on $v_n^{\rm inc}$ is 
independent for the two methods and mostly reflects the different purity
of $>$95\% compared to $>$99\% for the calorimeter and conversion method, 
respectively. 

\begin{table*}[htb]
\caption{
Summary of systematic uncertainties on the input to the measurement of 
$v_n^{\rm dir}$, where the $R_\gamma$ is from Ref.~\cite{ppg162}, and the 
$v_n^{\rm inc}$ and $v_n^{\rm dec}$ indicate ``inclusive" 
and ``decay" photons, respectively.
The values are quoted for \pt $<$ 3 \gevc, although most do not 
vary with \pt, as can be seen from Figures~\ref{fig:vincl} and 
\ref{fig:vpion}. The uncertainties on the $v_n^{\rm dec}$ due to the 
statistical uncertainty of the input data are uncorrelated between data 
points (type A); they are included in the statistical errors on the final 
results. Type B uncertainties are correlated in \pt, i.e. they can vary 
with \pt but only smoothly in the quoted range. Type C uncertainties 
change $v_n^{\rm dir}$ for all \pt by a constant multiplicative factor.  
The systematic uncertainties on \vtwo and \vthr are typical values.
}
\label{tab:syserr}
\begin{ruledtabular} \begin{tabular}{clcccc}
 &        & \multicolumn{3}{c}{centralities} & \\
Input & Source & 0\%--20\% & 20\%--40\% & 40\%--60\% & Type \\
\hline
$R_\gamma$ && 5.5\% & 5.5\% & 5.5\% & B \\
\\
$v^{\rm inc}_2$  
& conversion method   & $<$1\% & $<$1\% & $<$1\% & B \\
& calorimeter method  & 4\%    & 3\%    & 4\%    & B \\
\\
$v^{\rm dec}_2$
& meson $v_2$ (stat)  & $<$1\% & $<$1\% & $<$1\% & A \\
& $\pi^0$ $v_2$ (sys) & 5\%    & 3\%    & 2\%    & B \\
& $\eta,\omega$ $v_2$ (sys)  & $<$1\% & $<$1\% & $<$1\%  & B \\
& \bf{Event plane}         & 3\%    & 3\%    & 3\%    & C \\
\\
$v^{\rm inc}_3$  
&conversion method   & $<$1\% & $<$1\% & $<$1\% & B \\
&calorimeter method  & 5\%    & 7\%    & 10\%   & B \\
\\
$v^{\rm dec}_3$  
&meson $v_3$ (stat)  & 1\%    & 2\%    & 4\%    & A \\
&$\pi^0$ $v_3$ (sys)   & 11\%   & 11\%   & 11\%   & B \\
&$\eta,\omega$ $v_3$ (sys)  & $\sim$ 1\% & $\sim$1\% & $\sim$1\%  & B \\
& \bf{Event plane}         & 6\%    & 7\%   & 18\%   & C \\
\end{tabular} \end{ruledtabular}
\end{table*}

Using Gaussian error propagation, the statistical and systematic 
uncertainties would be calculated as:
\begin{eqnarray}
\sigma_{v_n^{\rm dir}}^2 
&=& \Big( \frac{R_\gamma}{R_\gamma-1} \Big)^2 
\times \sigma_{v_n^{\rm inc}}^{2}+ \Big( \frac{1} {R_{\gamma}-1} \Big)^2 \nonumber \\
&\times& \sigma_{v_n^{\rm dec}}^{2}+ \Big( \frac{v_n^{\rm dec}-v_n^{\rm inc}}{R_\gamma-1} \Big)^2 
\times \sigma_{R_\gamma}^{2}+ \sigma_{EP}^2.
\label{eq:syserr}
\end{eqnarray}
Except for the case $v_n^{\rm inc} = v_n^{\rm dec}$, there is a nonlinear 
dependence on $R_{\gamma}$ that, combined with uncertainties of 20\%--30\% on 
$(R_\gamma-1)$, results in asymmetric uncertainties, which are not 
described by Eq.~\ref{eq:syserr}. In particular, for the case 
$v_n^{\rm dec}>v_n^{\rm inc}$ the uncertainties on $v_n^{\rm dec}$ and 
$v_n^{\rm inc}$ are amplified if $R_{\gamma}$ is small.

\begin{figure}[htbp]
 \includegraphics[width=0.998\linewidth]{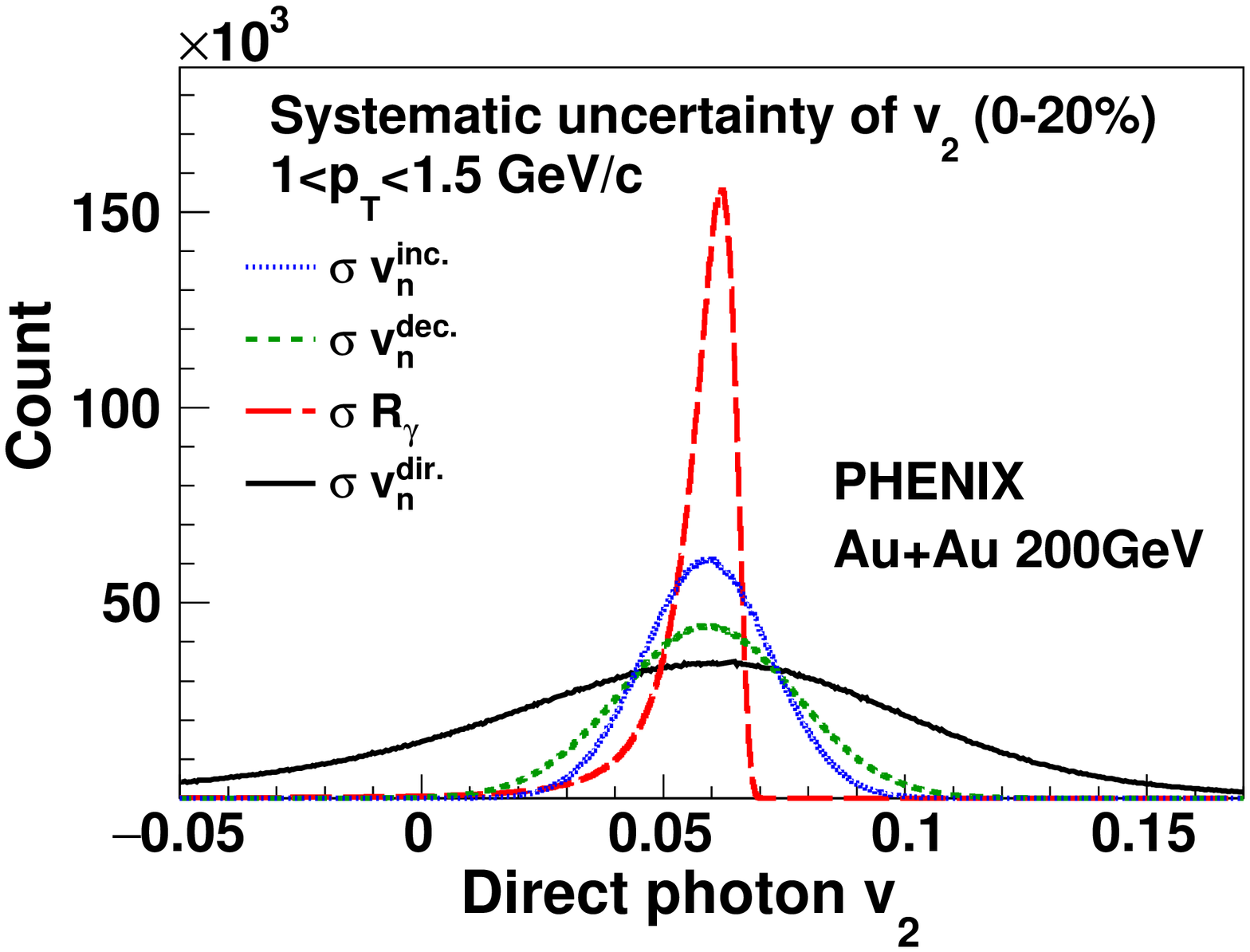}
  \caption{ 
This example shows the direct photon $v_{2}^{\rm dir}$ measured via the 
calorimeter method with the event plane estimated by the reaction plane 
detector (1 $< |\eta| <$ 2.8) in the 0\%--20\% centrality bin. Each of the 
various dashed curves indicate the probability distribution of the 
$v_{2}^{\rm dir}$ result due to the variation of a single term in 
Eq.~\ref{eq:vdirect}. While varying $v_2^{\rm inc}$ and $v_2^{\rm dec}$ 
alone 
leaves the uncertainty on $v_2^{\rm dir}$ Gaussian, varying $R_{\gamma}$ 
results in strongly asymmetric shapes.  The black solid curve shows the 
result when all uncertainties are taken into account simultaneously.
  }
    \label{fig:syserrmc}
\end{figure}

We estimate these asymmetric uncertainties by modeling a probability 
distribution for possible values of $v_n^{\rm dir}$ using the statistical and 
systematic uncertainties on $v_n^{\rm inc},v_n^{\rm dec},R_{\gamma}$, and the 
event plane resolution. We assume that the individual statistical and 
systematic uncertainties follow Gaussian probability distributions. The 
probability distribution for $v_n^{\rm dir}$ is then determined by generating 
many combinations of $v_n^{\rm inc}$, $v_n^{\rm dec}$, and $R_{\gamma}$. 
Figure~\ref{fig:syserrmc} shows one example of a probability distribution 
based on the systematic uncertainties on the calorimeter measurement for 
0\%--20\% centrality and $1<\pt<1.5$\,\gevc. In Figure~\ref{fig:syserrmc} 
the effect of the 
uncertainty of only $v_n^{\rm inc}, v_n^{\rm dec}$ or $R_{\gamma}$, are plotted 
separately. The asymmetry due to the uncertainty of $R_\gamma$ is clearly 
visible.

Probability distributions based on statistical (including type A 
systematics) and systematic uncertainties are determined for each 
$v_n^{\rm dir}$ data point in \pt and centrality and for both analyses. The 
central value for each data point was calculated using 
Eq.~\ref{eq:vdirect}.  We note that the peak or median of the 
probability distributions used to determine the statistical and systematic 
uncertainties agrees with the calculated central value to better than the 
symbol size. From each distribution we calculate the lower and upper bound 
on the uncertainty by integrating from $\pm\infty$ to a $v_n$ for which 
the integrated probability reaches 15.9\%. These values bracket a 68\% 
probability range for $v_n$ and are quoted as upper and lower statistical 
and systematic uncertainties on the final result.

\begin{figure*}[htbp]
  \includegraphics[width=1.0\linewidth]{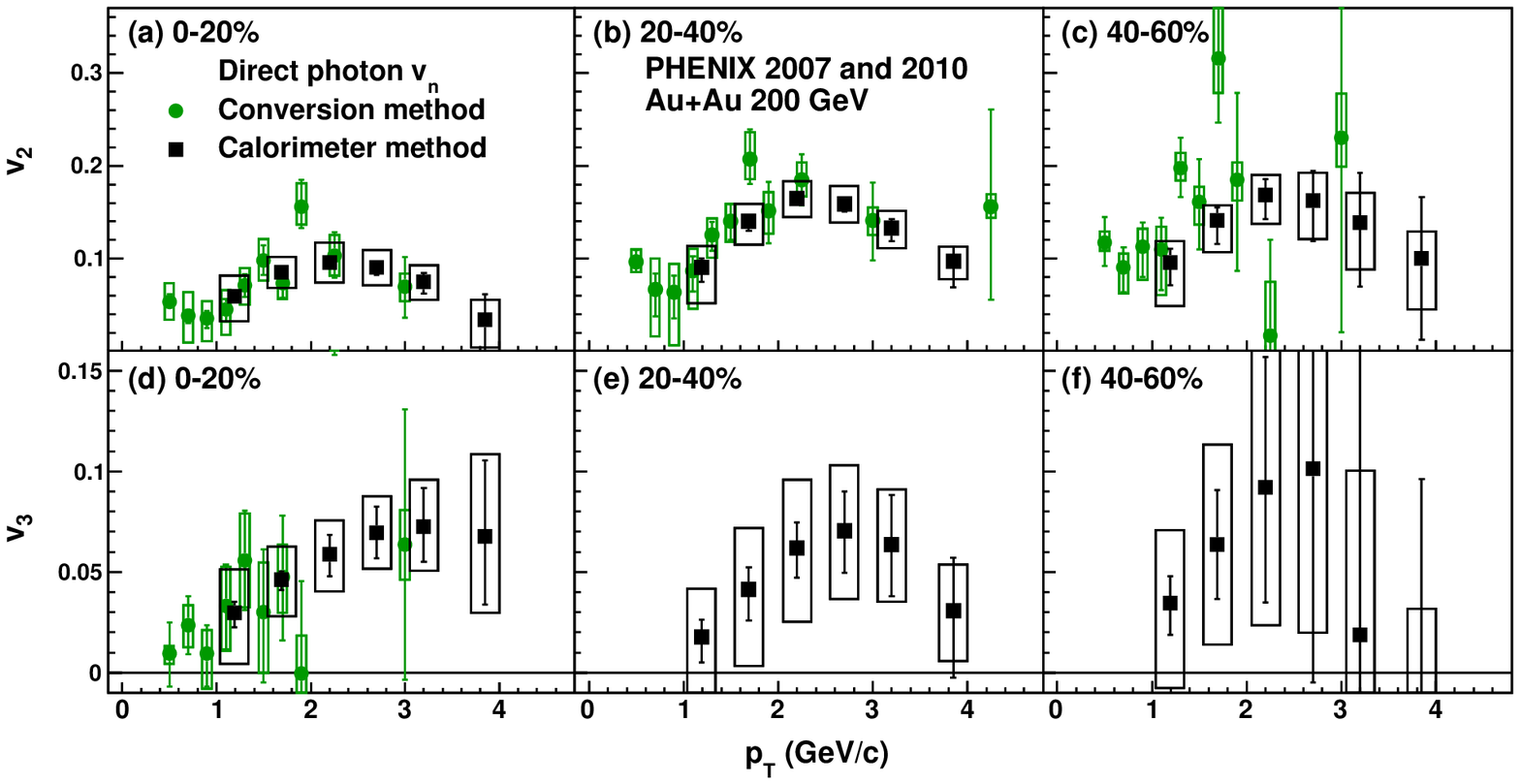}
  \caption{ 
Direct photon \vtwo and \vthr at midrapidity ($|\eta| <$ 0.35), for 
different centralities, measured with the conversion method (solid 
circles, green) and calorimeter method (solid squares, black).  The event 
plane was determined with the reaction plane detector (1 $< |\eta| <$ 
2.8). The error bars (boxes) around the data points are statistical 
(systematic) uncertainties.}
   \label{fig:vdir}
\end{figure*}

The final results for the direct photon \vtwo and \vthr, including 
statistical and systematic uncertainties as outlined above, are shown in 
Figure~\ref{fig:vdir} for three centralities and separately for the two 
analysis methods.  For the conversion method \vthr is shown only for the 
highest centrality bin; the statistical fluctuations preclude any 
meaningful measurement in the more peripheral bins. The data and their 
uncertainties are shown in Tables~\ref{tab:photon_v2} 
and~\ref{tab:photon_v3}.

\begingroup \squeezetable
\begin{table*}[htb]
  \caption{\label{tab:photon_v2}
Direct photon \vtwo for the indicated centrality  
bins for the two methods used.  Uncertainties are shown 
separately as upper and lower.
}
\begin{ruledtabular}
   \begin{tabular}{cccccc}
Centrality & Method &   $<\pt>$ [\gevc] & \vtwo & Statistical uncert. 
& Systematic uncert. \\ 
\hline
0\%--20\% & conversion photon &   
     0.50 &    0.0531 &    +0.0084,  -0.0076 &    +0.0200,  -0.0187 \\
&&   0.70 &    0.0387 &    +0.0070,  -0.0087 &    +0.0252,  -0.0291 \\
&&   0.90 &    0.0357 &    +0.0080,  -0.0104 &    +0.0185,  -0.0246 \\
&&   1.10 &    0.0456 &    +0.0105,  -0.0135 &    +0.0208,  -0.0277 \\
&&   1.30 &    0.0713 &    +0.0116,  -0.0128 &    +0.0185,  -0.0207 \\
&&   1.50 &    0.0979 &    +0.0162,  -0.0153 &    +0.0227,  -0.0214 \\
&&   1.70 &    0.0735 &    +0.0148,  -0.0160 &    +0.0157,  -0.0173 \\
&&   1.90 &    0.1560 &    +0.0291,  -0.0229 &    +0.0254,  -0.0192 \\
&&   2.25 &    0.1034 &    +0.0247,  -0.0243 &    +0.0223,  -0.0215 \\
&&   3.00 &    0.0699 &    +0.0316,  -0.0338 &    +0.0140,  -0.0155 \\
&&   4.25 &   -0.3534 &    +0.8077,  -0.1197 &    +0.1149,  -0.1831 \\
& calorimeter &   
     1.19 &    0.0591 &    +0.0038,  -0.0058 &    +0.0225,  -0.0266 \\
&&   1.69 &    0.0852 &    +0.0029,  -0.0035 &    +0.0163,  -0.0170 \\
&&   2.20 &    0.0957 &    +0.0046,  -0.0050 &    +0.0214,  -0.0218 \\
&&   2.70 &    0.0903 &    +0.0074,  -0.0078 &    +0.0186,  -0.0190 \\
&&   3.20 &    0.0747 &    +0.0098,  -0.0122 &    +0.0177,  -0.0189 \\
&&   3.85 &    0.0339 &    +0.0282,  -0.0430 &    +0.0218,  -0.0298 \\
20\%--40\% & conversion photon &   
     0.50 &    0.0964 &    +0.0125,  -0.0113 &    +0.0133,  -0.0113 \\
&&   0.70 &    0.0668 &    +0.0173,  -0.0289 &    +0.0336,  -0.0485 \\
&&   0.90 &    0.0640 &    +0.0178,  -0.0281 &    +0.0308,  -0.0555 \\
&&   1.10 &    0.0866 &    +0.0155,  -0.0217 &    +0.0240,  -0.0403 \\
&&   1.30 &    0.1251 &    +0.0146,  -0.0170 &    +0.0178,  -0.0240 \\
&&   1.50 &    0.1405 &    +0.0182,  -0.0202 &    +0.0185,  -0.0227 \\
&&   1.70 &    0.2074 &    +0.0316,  -0.0269 &    +0.0291,  -0.0212 \\
&&   1.90 &    0.1511 &    +0.0314,  -0.0342 &    +0.0207,  -0.0245 \\
&&   2.25 &    0.1846 &    +0.0279,  -0.0273 &    +0.0186,  -0.0174 \\
&&   3.00 &    0.1412 &    +0.0407,  -0.0431 &    +0.0137,  -0.0160 \\
&&   4.25 &    0.1561 &    +0.1048,  -0.0992 &    +0.0133,  -0.0121 \\ 
& calorimeter &
     1.19 &    0.0902 &    +0.0097,  -0.0151 &    +0.0236,  -0.0377 \\
&&   1.69 &    0.1403 &    +0.0066,  -0.0104 &    +0.0185,  -0.0248 \\
&&   2.20 &    0.1649 &    +0.0046,  -0.0056 &    +0.0188,  -0.0202 \\
&&   2.70 &    0.1592 &    +0.0071,  -0.0083 &    +0.0189,  -0.0200 \\
&&   3.20 &    0.1327 &    +0.0098,  -0.0136 &    +0.0190,  -0.0216 \\
&&   3.85 &    0.0972 &    +0.0155,  -0.0277 &    +0.0153,  -0.0192 \\
40\%--60\% & conversion photon & 
     0.50 &    0.1173 &    +0.0272,  -0.0252 &    +0.0117,  -0.0086 \\
&&   0.70 &    0.0905 &    +0.0214,  -0.0266 &    +0.0149,  -0.0280 \\
&&   0.90 &    0.1128 &    +0.0261,  -0.0327 &    +0.0192,  -0.0349 \\
&&   1.10 &    0.1101 &    +0.0338,  -0.0444 &    +0.0243,  -0.0473 \\
&&   1.30 &    0.1978 &    +0.0325,  -0.0313 &    +0.0163,  -0.0138 \\
&&   1.50 &    0.1608 &    +0.0465,  -0.0508 &    +0.0168,  -0.0244 \\
&&   1.70 &    0.3154 &    +0.0943,  -0.0687 &    +0.0771,  -0.0366 \\
&&   1.90 &    0.1848 &    +0.0943,  -0.0969 &    +0.0184,  -0.0224 \\
&&   2.25 &    0.0173 &    +0.1036,  -0.1478 &    +0.0584,  -0.1188 \\
&&   3.00 &    0.2305 &    +0.2262,  -0.1954 &    +0.0473,  -0.0310 \\
&&   4.25 &   -0.0043 &    +0.4198,  -0.2826 &    +0.0466,  -0.0920 \\
& calorimeter &
     1.19 &    0.0960 &    +0.0147,  -0.0247 &    +0.0226,  -0.0462 \\
&&   1.69 &    0.1412 &    +0.0139,  -0.0255 &    +0.0162,  -0.0334 \\
&&   2.20 &    0.1687 &    +0.0172,  -0.0258 &    +0.0212,  -0.0313 \\
&&   2.70 &    0.1624 &    +0.0323,  -0.0427 &    +0.0302,  -0.0405 \\
&&   3.20 &    0.1388 &    +0.0539,  -0.0657 &    +0.0319,  -0.0487 \\
&&   3.85 &    0.0999 &    +0.0670,  -0.0788 &    +0.0290,  -0.0533 \\
   \end{tabular}
   \end{ruledtabular}
\end{table*}
\endgroup

\begingroup \squeezetable
\begin{table*}[!ht]
  \caption{
Direct photon \vthr for the indicated centrality  
bins for the two methods used.  Uncertainties are shown 
separately as upper and lower.
}
\label{tab:photon_v3}
\begin{ruledtabular}
   \begin{tabular}{cccccc}
Centrality & Method &   $<\pt>$ [\gevc] & \vthr & Statistical uncert. 
& Systematic uncert. \\ 
\hline
0\%--20\% & conversion photon &   
     0.50 &    0.0094 &    +0.0155,  -0.0163 &    +0.0039,  -0.0052 \\
& &  0.70 &    0.0237 &    +0.0142,  -0.0146 &    +0.0099,  -0.0111 \\
& &  0.90 &    0.0094 &    +0.0143,  -0.0163 &    +0.0119,  -0.0173 \\
& &  1.10 &    0.0333 &    +0.0204,  -0.0218 &    +0.0193,  -0.0223 \\
& &  1.30 &    0.0558 &    +0.0247,  -0.0247 &    +0.0233,  -0.0233 \\
& &  1.50 &    0.0299 &    +0.0314,  -0.0346 &    +0.0246,  -0.0301 \\
& &  1.70 &    0.0476 &    +0.0305,  -0.0317 &    +0.0161,  -0.0177 \\
& &  1.90 &   -0.0006 &    +0.0461,  -0.0535 &    +0.0189,  -0.0265 \\
& &  2.25 &    0.2094 &    +0.0657,  -0.0516 &    +0.0461,  -0.0299 \\
& &  3.00 &    0.0637 &    +0.0672,  -0.0672 &    +0.0172,  -0.0174 \\
& &  4.25 &    0.2753 &    +0.4140,  -0.4118 &    +0.1492,  -0.0765 \\ 
& calorimeter &
     1.19 &    0.0298 &    +0.0055,  -0.0073 &    +0.0214,  -0.0256 \\
&&   1.69 &    0.0461 &    +0.0040,  -0.0053 &    +0.0166,  -0.0182 \\
&&   2.20 &    0.0587 &    +0.0096,  -0.0110 &    +0.0170,  -0.0185 \\
&&   2.70 &    0.0696 &    +0.0129,  -0.0129 &    +0.0180,  -0.0180 \\
&&   3.20 &    0.0726 &    +0.0191,  -0.0175 &    +0.0231,  -0.0221 \\
&&   3.85 &    0.0677 &    +0.0380,  -0.0332 &    +0.0408,  -0.0378 \\
20\%--40\% & calorimeter &
     1.19 &    0.0178 &    +0.0085,  -0.0127 &    +0.0240,  -0.0343 \\
&&   1.69 &    0.0415 &    +0.0108,  -0.0154 &    +0.0304,  -0.0381 \\
&&   2.20 &    0.0619 &    +0.0128,  -0.0146 &    +0.0339,  -0.0365 \\
&&   2.70 &    0.0703 &    +0.0198,  -0.0206 &    +0.0326,  -0.0336 \\
&&   3.20 &    0.0637 &    +0.0244,  -0.0256 &    +0.0274,  -0.0284 \\
&&   3.85 &    0.0308 &    +0.0265,  -0.0331 &    +0.0228,  -0.0250 \\
40\%--60\% & calorimeter &
     1.19 &    0.0346 &    +0.0131,  -0.0157 &    +0.0362,  -0.0422 \\
&&   1.69 &    0.0638 &    +0.0271,  -0.0273 &    +0.0497,  -0.0494 \\
&&   2.20 &    0.0920 &    +0.0651,  -0.0567 &    +0.0780,  -0.0676 \\
&&   2.70 &    0.1011 &    +0.1224,  -0.1028 &    +0.0973,  -0.0793 \\
&&   3.20 &    0.0187 &    +0.1580,  -0.1476 &    +0.0823,  -0.0877 \\
&&   3.85 &   -0.0430 &    +0.1421,  -0.1289 &    +0.0751,  -0.0938 \\
   \end{tabular}
   \end{ruledtabular}
\end{table*}
\endgroup

The two analysis techniques are very different but the results agree well 
in the overlap region, and they are also consistent with the results 
published earlier~\cite{ppg126}. The direct photon \vtwo centrality 
dependence, both in trend and magnitude, is quite similar to the observed 
pion \vtwo. The third order coefficients \vthr are consistent with no 
centrality dependence.

		\section{Comparisons to models}
		\label{sec:modelcomp}

As already mentioned, the essence of the ``direct photon puzzle'' is that 
current theoretical scenarios have difficulties explaining the large 
direct photon yield and azimuthal asymmetries at the same time. This is 
illustrated by a recent state-of-the-art calculation of viscous hydrodynamic calculation of photon emission with fluctuating initial density profiles and standard thermal 
rates \cite{shen:2015visc}, which falls significantly short in describing 
yield and \vtwo. Over the past few years many new ideas have been proposed 
to resolve this puzzle, including non equilibrium effects 
\cite{monnai:2014scf, 
gelis:2004noneq,mclerran:2015tail,mclerran:2014glas}, enhanced early 
emission due to large magnetic fields 
\cite{Muller:2013ila,basar:2012conf,tuchin:2013magn}, enhanced emission at 
hadronization \cite{campbell}, as well as modifications of the formation 
time and initial conditions 
\cite{chatterjee:2013flow,fmliu:2014ft,vujanovic:2014early}.

In this subsection we compare our results to a subset of the models which 
(i) consider thermal radiation from the QGP and HG (hadron gas) plus 
additional proposed sources, (ii) have a complete model for the space-time 
evolution, and (iii) calculate absolute yields and \vtwo. For the 
comparison we use the data for the 20\%--40\% centrality class, and note that 
the comparison leads to similar conclusions for the other centrality bins. 
While none of the models describes all aspects of the available data, they 
are representative of how different theories are trying to cope with the 
challenge.

\begin{figure}[htbp]
  \includegraphics[width=0.998\linewidth]{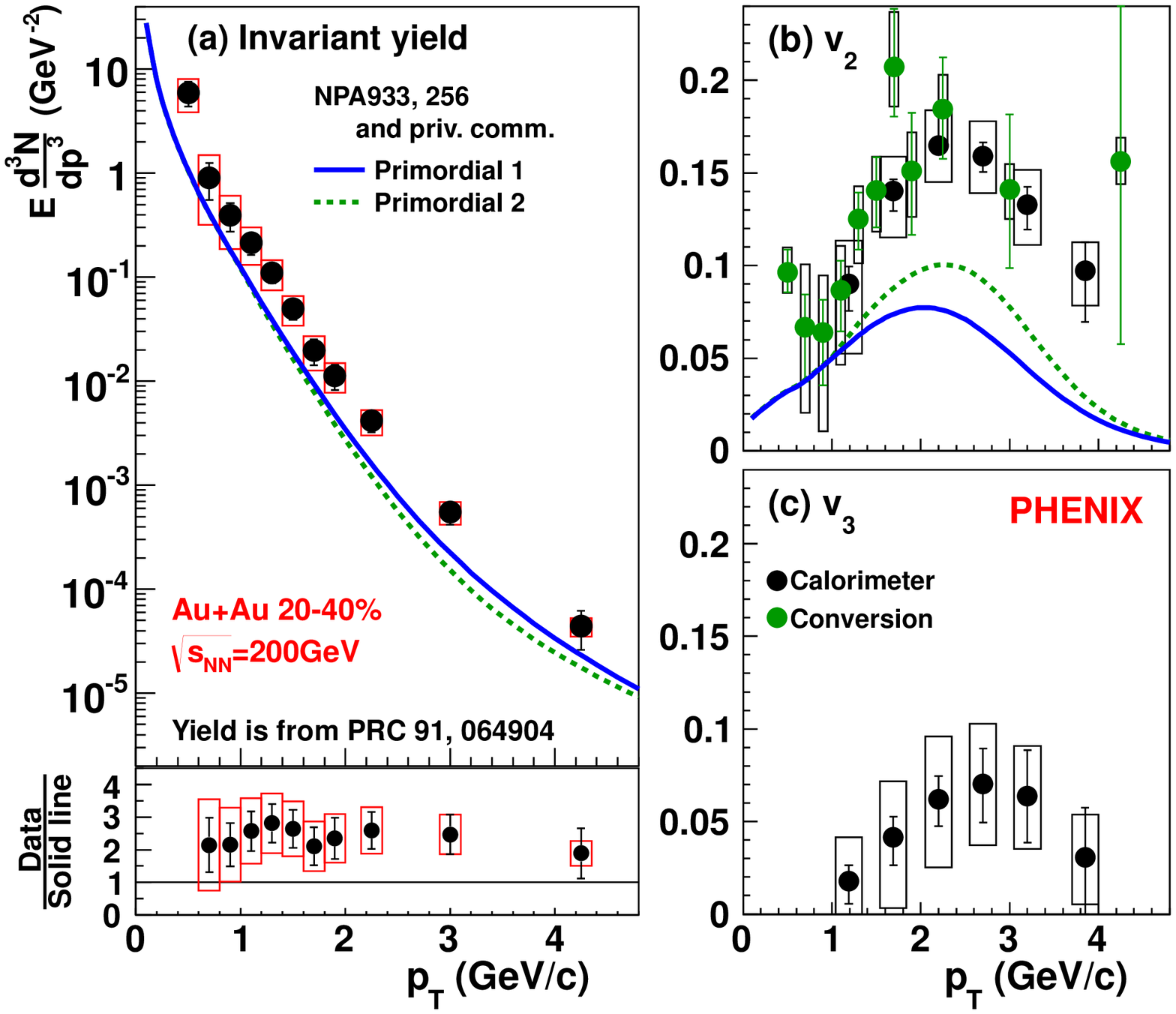}
  \caption{Comparison of the direct photon yields~\cite{ppg162} 
and \vtwo with the fireball model~\cite{hees:2015pce}. 
The two curves for \vtwo correspond to two different 
parametrizations of the prompt photon component. See text for details.
  }
    \label{fig:TheoryOpt1}
\end{figure}

First, we compare the data to the ``fireball'' scenario originally 
calculated in Ref.~\cite{hees:2011fb}. The model includes pQCD, QGP and HG 
contributions, with the instantaneous rates convoluted with a fireball 
expansion profile. The basic parameter is the initial transverse 
acceleration of the fireball, $a_T$. The prompt photon component is 
estimated in two ways. The first variant is a parametrization of the 
photon yields measured in \pp by the PHENIX experiment~\cite{ppg060} 
(labeled as ``primordial 1"), the second is an $x_t$-scaling motivated 
parametrization (labeled as ``primordial 2"), modified with an empirical 
factor $K=2.5$ to match the measured data at high \pt (above 4\,\gevc). 
The yield calculation includes thermal yields from the QGP with 
$T_0=350$\,MeV and from the hadronic phase.  Different from an earlier 
version of the model, chemical equilibrium prior to kinetic freeze-out is 
no longer assumed.  This results in a large enhancement in photon 
production in the later hadronic stages via processes like meson 
annihilation (for instance $\pi+\rho\rightarrow\pi+\gamma$). With an 
initial transverse acceleration $a_T=0.12$\,$c^2$/fm and 
$\tau\approx15$\,fm/$c$ fireball lifetime, 100\,MeV freeze-out temperature 
and $\beta_s=0.77$ surface velocity, the observed low \pt photon yields 
are recovered within systematic uncertainties, but underpredict the 
data~\cite{hees:2011fb}. In Figure~\ref{fig:TheoryOpt1} the data are 
compared to the most recent updated ``fireball'' scenario shown in 
Ref.~\cite{hees:2015pce}, which includes a calculation with ideal 
hydrodynamics with finite initial flow at thermalization and enhanced 
yields around chemical freeze-out temperature $T_c$ that improves the 
description of the data. The direct photon \vtwo has its maximum at about 
the same \pt in both theory and data. The \vtwo calculated in the original 
fireball scenario~\cite{hees:2011fb} under predicts the measured one. The 
radial boost hardens the photons from the hadronic gas (HG) and in this 
way increases \vtwo as well, but the calculation still falls short of the 
measurement. \vthr is currently not calculated in this model.

\begin{figure}[htbp]
  \includegraphics[width=0.998\linewidth]{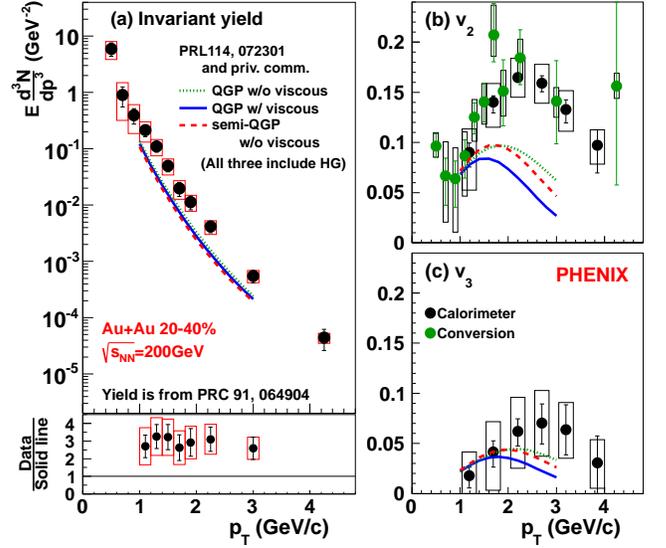}
    \caption{ 
Comparison of the direct photon yields and \vtwo, \vthr with a 
hydrodynamical model~\cite{Ryu:2015vwa,Paquet:2015lta} calculated under 
three different assumptions including the ``semi-QGP'' scenario~\cite{gale:2015semi}.
      }
    \label{fig:TheoryOpt0}
\end{figure}

Second, in Figure~\ref{fig:TheoryOpt0} the data are compared to three 
calculations evaluated with the hydrodynamical background as described in 
Ref.~\cite{Ryu:2015vwa,Paquet:2015lta}. The first calculation, labeled 
"QGP w/ viscous", was evaluated using the AMY photon emission rate in the high-temperature 
(QGP) region, and included viscous corrections to the photon emission 
rates~\cite{Shen:2014nfa,dion:2011visc} due to both bulk and shear 
viscosities. The same calculation without the viscous corrections 
corresponds to the curve labeled "QGP, w/o viscous".  Once viscous 
corrections are included, \vtwo drops by more than 50\% at 3\,\gevc, while 
the yield decreases just by $\sim$10\%. The third curve, labelled 
"semi-QGP, w/o viscous", shows the consequence of including the effect of 
confinement on the photon emission rate, as computed in the semi-QGP 
approach~\cite{gale:2015semi}. The utilization of the semi-QGP photon 
rates at high temperatures suppresses the spectrum, but does not change 
the $v_n$ significantly. This is a consequence of the small contribution 
of QGP photons to the thermal photon \vtwo, which is dominantly produced 
at temperatures around and smaller than the confinement temperature. The 
prompt photon contributions in all three calculations are evaluated within 
the perturbative QCD framework.

\begin{figure}[htbp]
  \includegraphics[width=0.998\linewidth]{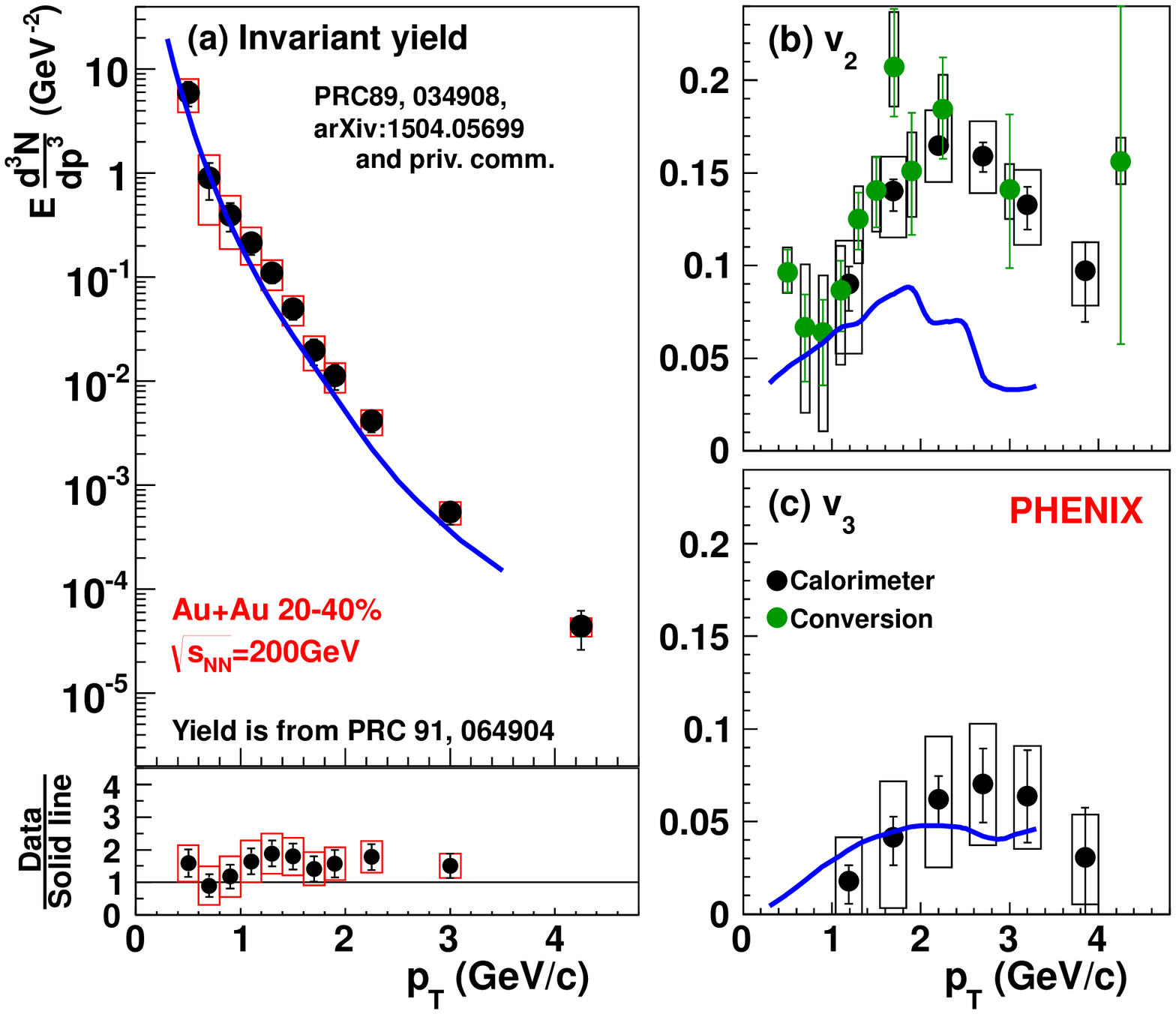}
    \caption{  
Comparison of the direct photon yields and \vtwo with the PHSD 
model~\cite{linnyk:phsd2014,linnyk:phsd2015}.
      }
    \label{fig:TheoryOpt2}
\end{figure}

Third, we compare the data with PHSD (parton-hadron-string dynamics), a 
microscopic transport model~\cite{linnyk:phsd2014}. In addition to the 
traditional QGP and HG sources (resonance decays) this model includes late 
stage meson-meson and meson-baryon Bremsstrahlung, which enhances the 
yield at the lowest \pt substantially and increases \vtwo by almost 50\% 
in the \pt$<3$\,\gevc region (see Figure 2 in 
Ref.~\cite{linnyk:phsd2014}).  Contributions from photonic decays of 
$\phi$ and $a_1$ are also included, because these are not subtracted in the 
measurement.  After all other sources are added, the direct photon 
spectrum is very well reproduced below 3\,\gevc, but \vtwo under predicts 
the measured values. Also, the \pt where \vtwo reaches its maximum is 
under predicted. In Figure~\ref{fig:TheoryOpt2} the data are compared to 
the latest PHSD model calculation~\cite{linnyk:phsd2015} that included 
additional photon production channels in the hadronic phase and improved 
the Bremsstrahlung calculation.  The model also provides \vthr. It is 
positive and consistent with the data within uncertainties.

Explaining the large yield and strong flow simultaneously requires 
significant improvements in quantifying the contributions from the late 
stage QGP and hadron-gas interactions. Even deeper insight on both the 
photon sources and the time profile of the system may be necessary to 
further improve the models. Future measurements of more differential 
quantities will help to distinguish and quantify the individual photon 
sources.


		\section{Summary and conclusions}
		\label{sec:summary}

The PHENIX experiment at the Relativistic Heavy Ion Collider measured 2nd 
and 3rd order Fourier coefficients of the azimuthal distributions of 
direct photons emitted at midrapidity in \sqsn = 200\,GeV \auau 
collisions, for various collision centralities.  Two different and 
independent analyses are used to determine the inclusive photon yield. The 
external conversion photon measurement allows one to extend the \pt range 
down to 0.4 \gevc compared to 1.0 \gevc for the calorimetric measurement. 
In the overlap region the two results are consistent.  The \vtwo 
measurements are also consistent with earlier published results, while 
\vthr is published for the first time.

Both the direct photon \vtwo and \vthr are found to be large. The \vtwo 
exhibits a clear centrality-dependence, while \vthr is consistent with no 
centrality dependence. At all centralities, the direct photon \vtwo is 
similar in magnitude to the hadron \vtwo for \pt $<$ 3\,\gevc, The direct 
photon \vthr is consistent with that for hadrons over the entire \pt 
range.

We compare the data to several recent calculations, which treat the direct 
photon yields and the azimuthal asymmetries in a consistent production and 
evolution framework. None of them describe the full systematics of the 
data adequately, but there has been progress in the last few years. The 
general trend of the models appears to be including sources from the 
earliest (pre-equilibrium, see for instance Ref.~\cite{Muller:2013ila}) 
or very late times in the evolution of the system, while giving less 
emphasis to photon production at intermediate times, when most of the 
expansion occurs. PHSD includes new sources from the hadron gas and photon 
production even after the hadrons are decoupled from each other, which 
improves description of the yields but still under predicts \vtwo.  The 
model that best approximates the measured \vtwo, including the \pt region 
where \vtwo reaches its maximum value, starts the evolution with a large 
initial boost even before thermalization~\cite{hees:2011fb}.  It is also 
worth noting that the microscopic transport model~\cite{linnyk:phsd2014} 
is able to describe the anisotropies as well as the full-scale viscous 
hydrodynamics~\cite{gale:2015semi}.

While the data are getting more differential and more accurate, and model 
calculations improve, the ``direct photon puzzle'' remains unresolved. 
High quality data of yields and \vtwo and \vthr for different collision 
systems, including very asymmetric ones, and energies would help to 
further improve our understanding of direct photon production because robust 
models must be able to describe the data over a wide range of experimental 
conditions.


		\section*{ACKNOWLEDGMENTS}   

We thank the staff of the Collider-Accelerator and Physics
Departments at Brookhaven National Laboratory and the staff of
the other PHENIX participating institutions for their vital
contributions.  We acknowledge support from the 
Office of Nuclear Physics in the
Office of Science of the Department of Energy,
the National Science Foundation, 
Abilene Christian University Research Council, 
Research Foundation of SUNY, and
Dean of the College of Arts and Sciences, Vanderbilt University 
(U.S.A),
Ministry of Education, Culture, Sports, Science, and Technology
and the Japan Society for the Promotion of Science (Japan),
Conselho Nacional de Desenvolvimento Cient\'{\i}fico e
Tecnol{\'o}gico and Funda\c c{\~a}o de Amparo {\`a} Pesquisa do
Estado de S{\~a}o Paulo (Brazil),
Natural Science Foundation of China (Peoples' Republic of~China),
Croatian Science Foundation and
Ministry of Science, Education, and Sports (Croatia),
Ministry of Education, Youth and Sports (Czech Republic),
Centre National de la Recherche Scientifique, Commissariat
{\`a} l'{\'E}nergie Atomique, and Institut National de Physique
Nucl{\'e}aire et de Physique des Particules (France),
Bundesministerium f\"ur Bildung und Forschung, Deutscher
Akademischer Austausch Dienst, and Alexander von Humboldt Stiftung (Germany),
National Science Fund, OTKA, K\'aroly R\'obert University College, 
and the Ch. Simonyi Fund (Hungary),
Department of Atomic Energy and Department of Science and Technology (India), 
Israel Science Foundation (Israel), 
Basic Science Research Program through NRF of the Ministry of Education (Korea),
Physics Department, Lahore University of Management Sciences (Pakistan),
Ministry of Education and Science, Russian Academy of Sciences,
Federal Agency of Atomic Energy (Russia),
VR and Wallenberg Foundation (Sweden), 
the U.S. Civilian Research and Development Foundation for the
Independent States of the Former Soviet Union, 
the Hungarian American Enterprise Scholarship Fund,
and the US-Israel Binational Science Foundation.


%
 
\end{document}